\documentclass[aps,prc,twocolumn,showpacs,showkeys,superscriptaddress]{revtex4-1}
\usepackage{longtable}
\usepackage{graphicx}
\usepackage{dcolumn}
\begin{document}
\title{Low lying spectroscopy of odd-odd $^{146}$Eu}
\author{T.~Bhattacharjee}
\thanks{Corresponding author}
\email{btumpa@vecc.gov.in}
\affiliation{Variable Energy Cyclotron Centre, Kolkata - 700064, India}
\author{D.~Banerjee}
\affiliation{Raiochemistry Division (BARC), Variable Energy Cyclotron Centre, Kolkata - 700 064, India}
\author{S.~K.~Das}
\affiliation{Raiochemistry Division (BARC), Variable Energy Cyclotron Centre, Kolkata - 700 064, India}
\author{S.~Chanda}
\affiliation{Fakir Chand College, Diamond Harbour, West Bengal,India}
\author{T.~Malik}
\affiliation{Department of Applied Physics, Indian School of Mines, Dhanbad, India}
\author{A.~Chowdhury}
\affiliation{Variable Energy Cyclotron Centre, Kolkata - 700064, India}
\author{P.~Das}
\affiliation{Variable Energy Cyclotron Centre, Kolkata - 700064, India}
\author{S.~Bhattacharyya}
\affiliation{Variable Energy Cyclotron Centre, Kolkata - 700064, India}
\author{R.~Guin}
\affiliation{Raiochemistry Division (BARC), Variable Energy Cyclotron Centre, Kolkata - 700 064, India}
\date{\today}
\begin{abstract}

Electron Capture (EC) decay of $^{146}$Gd($\it{t _{\frac{1}{2}}}$ = 48d) to the low lying states of $^{146}$Eu has been studied using high-resolution $\gamma$ ray spectroscopy. The $^{146}$Gd activity was produced by ($\alpha$, 2n) reaction at E$_{\alpha}$ = 32 MeV using 93.8$\%$ enriched $^{144}$Sm target. The level structure has been considerably modified from the measurement of $\gamma$ ray singles, $\gamma\gamma$ coincidences and  decay half lives. Lifetime measurement has been performed for the 3$^-$ (114.06 keV) and 2$^-$ (229.4 keV) levels of $^{146}$Eu using Mirror Symmetric Centroid Difference (MSCD) method with LaBr$_3$ (Ce) detectors. The lifetimes for these two states have been found to be 5.38 $\pm$ 2.36 ps and 8.38 $\pm$ 2.19 ps respectively. Shell model calculation has been performed using OXBASH code in order to interpret the results. \\

\end{abstract}



\maketitle

\section{Introduction}
\label{intro}
The low lying spectroscopy of nuclei, from both decay as well as in beam measurements~\cite{anagha,liddick,garret}, has drawn considerable attention in recent years. Such measurements for the odd-odd transitional nuclei in A $\sim$ 140 region are crucial in understanding the role of neutron proton interaction in the N, Z = 50 - 82 subshell space. The nuclei in the vicinity of the magic N = 82 and semi-magic Z = 64 exhibit excitations due to multiparticle-hole as well as quasi-vibrational
structures. The coupling of valence particles to the core phonon gives rise to complex (multi-)particle
phonon level families in these nuclei~\cite{kleinheinz}. The particle
(hole) interactions  and the resulting competition of the single particle with the underlying collective excitations
have been rigorously explored through a number of experimental and theoretical investigations~\cite{kader,podolyak,pm-sarmi,la136}.\\

Although the low lying levels in this mass region have been studied from decay and in-beam experiments with light ion beams, the information on level lifetimes and transition moments has been very rare in most of the cases. Moreover, these studies have been performed mostly with either NaI(Tl) or BaF$_2$ scintillators and intrinsic Ge(Li) detectors with limited time resolution.
Thus, it is important to undertake systematic measurements on lifetime and transition moments for many of these nuclei.
The measurement of lifetimes of the order of several ones or tens of picoseconds requires a detection system having a very good time resolution as well as a modest energy resolution. The precise technique for the measurements of level lifetimes of the order of few picoseconds has been a topic of research for many years~\cite{mach1, mach2}. With the availability of efficient inorganic scintillators, such as, LaBr$_3$(Ce), such measurements have taken a new direction~\cite{regis,ce138-new}.\\

The odd-odd $^{146}$Eu, with one extra neutron outside N = 82 shell closure is a topic of interest in order to test the persistence of subshell closure at Z = 64~\cite{gd146}. The structure of  $^{146}$Eu has been studied both from decay as well as in-beam reaction experiments~\cite{nndc,holmberg,kantus,ercan1,ercan2,b10,trans1}. The decay measurements have been carried out by different groups~\cite{nndc} in order to develop the excited levels as well as to obtain their lifetimes. Of these, R. Kantus et al.~\cite{kantus} have performed coincidence measurements with a Ge(Li) and a Gam-X detector and developed a decay scheme up to 690.7 keV excitation. However, according to the latest compilation in Nuclear Data Sheets~\cite{nndc}, the different measurements on $^{146}$Gd EC decay exhibit a lot of discrepancy in the observed levels, the energy and the intensity of the decay $\gamma$ rays.
The suggested sequence of the low lying states has been accounted for by combining a d$_{5/2}$ proton hole and a f$_{7/2}$ neutron particle following the jj coupling scheme of odd-odd nuclei.
The prompt decay of 9$^+$, 235 $\mu$s isomeric state of $^{146}$Eu has been rigorously studied by Ercan et al.~\cite{ercan1}.
The lifetime of the first 3$^-$ and 2$^-$ states has been measured by L. Holmberg et al.~\cite{holmberg} by using electron-electron coincidence measurement with a magnetic $\beta$ spectrometer where they could only provide the limits of the lifetime values as $\le$0.16ns and $\le$0.2ns for the two states respectively. However, the sequence of 114 and 115 keV $\gamma$ transitions has been modified afterwards by Kantus et al.~\cite{kantus}. Also, there exists a large variance (0.16 ns to 0.8 ns) in the measured lifetime of the 3$^-$ state~\cite{nndc}. Besides, the lifetime of the above two excited states seems to be quite large considering the suggested configuration of these states.\\

In the present work, we have reported the low lying level scheme of odd-odd $^{146}$Eu nucleus from the EC decay of 48d $^{146}$Gd. The  decay scheme of $^{146}$Gd has been modified considerably following the measurements of $\gamma$ ray singles, $\gamma\gamma$ coincidences and the decay half lives using high resolution Ge detectors. The lifetime of the first 3$^-$ and 2$^-$ states have been measured with LaBr$_3$ (Ce) detectors by employing the Mirror Symmetric  Centroid Difference (MSCD) method~\cite{regis}.
A shell model calculation has also been performed using OXBASH code~\cite{BAB94} in order to interpret the experimental results.

\section{Experimental Details}
\label{expt}
The excited states of $^{146}$Eu have been populated from the EC decay of $^{146}$Gd (t$_{1/2}$ = 48 days), produced via $^{144}$Sm ($\alpha$, 2n) reaction with 32 MeV alpha beam from K=130 AVF Cyclotron at Variable Energy Cyclotron Centre, Kolkata.
The $^{144}$Sm targets with a thickness of 300 $\mu$g/cm$^2$ were prepared on 6.84 mg/cm$^2$ Al foils by electro-deposition method using 93.8$\%$ enriched Sm$_2$O$_3$. The enriched sample contained $^{147}$Sm (2.06$\%$), $^{148}$Sm (1.00$\%$), $^{149}$Sm (0.95$\%$), $^{150}$Sm (0.41$\%$), $^{152}$Sm (1.05$\%$) and $^{154}$Sm (0.73$\%$) isotopes as impurity.
The theoretical cross sections for different nuclei produced in the reaction were estimated by using the code PACEIV~\cite{gavron}. The $^{146}$Gd nuclei produced from the above reaction were recoil-implanted on Al catcher foils (6.84 mg/cm$^2$) for the subsequent measurements. The Al catcher foils containing $^{146}$Gd activity were dissolved in acid solution which was subsequently used for $\gamma\gamma$ coincidence and lifetime measurements. The irradiated target was counted on a 50$\%$  HPGe detector for the measurements of $\gamma$ singles and decay half lives. In this measurement, the target was kept at a distance of 15 cm from the detector to ensure a dead time less than 10$\%$ and hence to minimise the summing effect.
A Canberra Digital System was used for biasing, pulse processing and data collection of the 50$\%$ HPGe detector.
The $\gamma$$\gamma$ coincidence data were acquired with a setup (called as `Ge-setup' later on in this paper) consisting of one 10$\%$ single HPGe detector and a segmented Low Energy Photon Spectrometer (LEPS), kept at an angle of 180$^{\circ}$ with respect to each other.  Another setup (called as `LaBr$_3$-setup' later on in this paper) consisting of two 30 mm x 30 mm LaBr$_3$ (Ce) detectors was used for the measurement of lifetimes.
In this setup, one detector was used as `START' detector and the other detector, kept at 180$^o$ with respect to the `START' detector, was used as the `STOP' detector. Time to Amplitude (TAC) signal was generated from these two detectors for the subsequent measurements.
In both the coincidence setups, used for $\gamma\gamma$ coincidence (Ge-setup) and lifetime measurements (LaBr$_3$-setup), 8K ADC and CAMAC based data acquisition system were used for collecting the zero suppressed list mode data with LAMPS~\cite{lamps} software.
The absolute efficiency of all the HPGe and LEPS detectors were determined by using calibrated $^{152}$Eu and $^{133}$Ba sources.

\section{Data Analysis and Results}
\label{da}

The data analysis has been performed for the measurements of (i) decay half lives with the 50$\%$ HPGe detector, (ii) $\gamma\gamma$ coincidence using Ge-setup and (iii) lifetime using LaBr$_3$-setup. The detailed procedures have been discussed in the following subsections along with the obtained results.

\subsection{Decay measurements and $\gamma\gamma$ coincidence measurements}
\label{decay-gg}
\begin{figure}
  \begin{center}
  \vskip 0.5cm
  \hskip -2.0cm
  % Requires \usepackage{graphicx}
  \includegraphics[height=0.8\columnwidth, width=9cm, angle=-90]{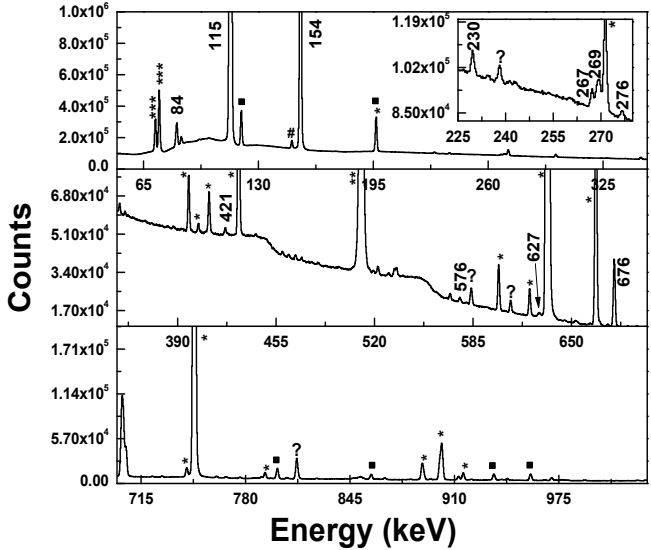}
  \vskip -2.0cm
  \caption{Total spectrum upto the EC decay Q value of $^{146}$Gd using 50$\%$ HPGe detector. $\gamma$ rays of $^{146}$Eu have been marked with their respective energy values. Peaks marked with $\star$, filled square and hash belong to the decay of $^{146}$Eu, $^{147}$Eu and $^{149}$Gd respectively. Annihilation $\gamma$ ray and some of the background transitions are marked with $\star \star$ and $\star \star \star$ respectively. The four peaks marked with question mark could not be identified.}
  \label{total-decay}
  \end{center}
\end{figure}

\begin{figure}
  \begin{center}
  \vskip -0.02cm
  \hskip -0.8cm
  % Requires \usepackage{graphicx}
  \includegraphics[height=0.8\columnwidth, width=9cm, angle=-90]{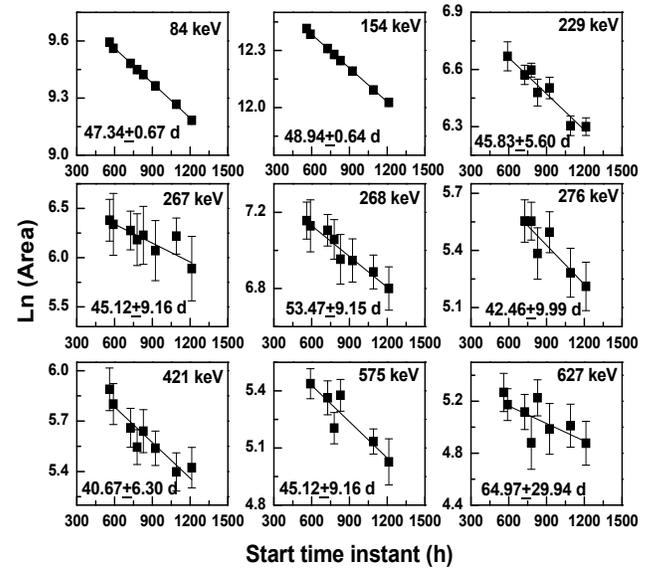}
  \vskip -1.4cm
  \caption{The decay plots of the $\gamma$ transitions corresponding to EC decay of $^{146}$Gd. The $\gamma$ energy and the half life value are indicated inside each figure.}
  \label{decay-all}
  \end{center}
\end{figure}

\begin{figure}
  \begin{center}
  \hskip -2.5cm
  % Requires \usepackage{graphicx}
  \includegraphics[height=0.7\columnwidth, width= 9cm, angle=-90]{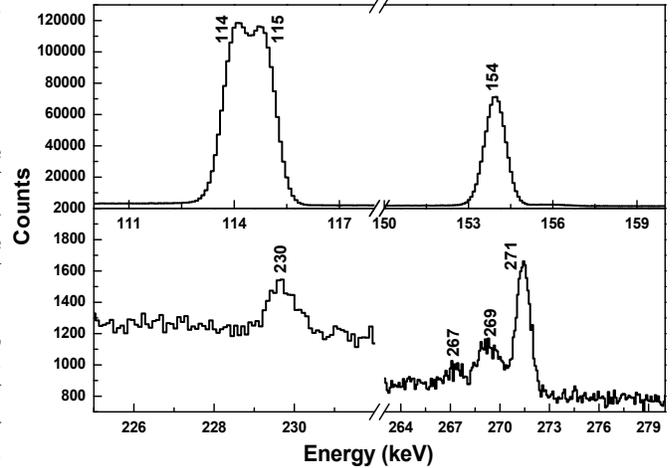}
  \vskip -2.5cm
  \caption{The total spectrum obtained with one segment of the LEPS detector. The separation of 114.06 keV and 114.88 keV is shown at the top panel and the presence of the 229.4, 267.02 and 268.96 keV transitions is shown in the bottom panel.}
  \label{leps-seg1}
  \end{center}
  \end{figure}

\begin{figure}
  \begin{center}
  \vskip 0.3cm
  \hskip -2.7cm
  % Requires \usepackage{graphicx}
  \includegraphics[height=0.75\columnwidth, width=9cm, angle=-90]{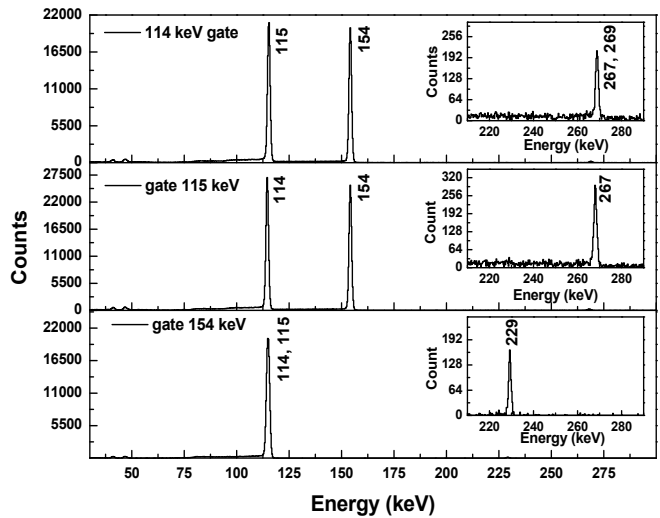}
  \vskip -2.3cm
  \caption{The gated spectra obtained from the 10$\%$ HPGe detector by putting gate on 114.06 keV (top), 114.88 keV (middle) and 153.86 keV(bottom). The insets show the region from 210 keV to 290 keV in an expanded scale.}
  \label{gate-all}
  \end{center}
\end{figure}

\begin{figure}
  \begin{center}
  \vskip 0.3cm
  \hskip -2.5cm
  % Requires \usepackage{graphicx}
  \includegraphics[height=0.75\columnwidth, width=8.5cm, angle=-90]{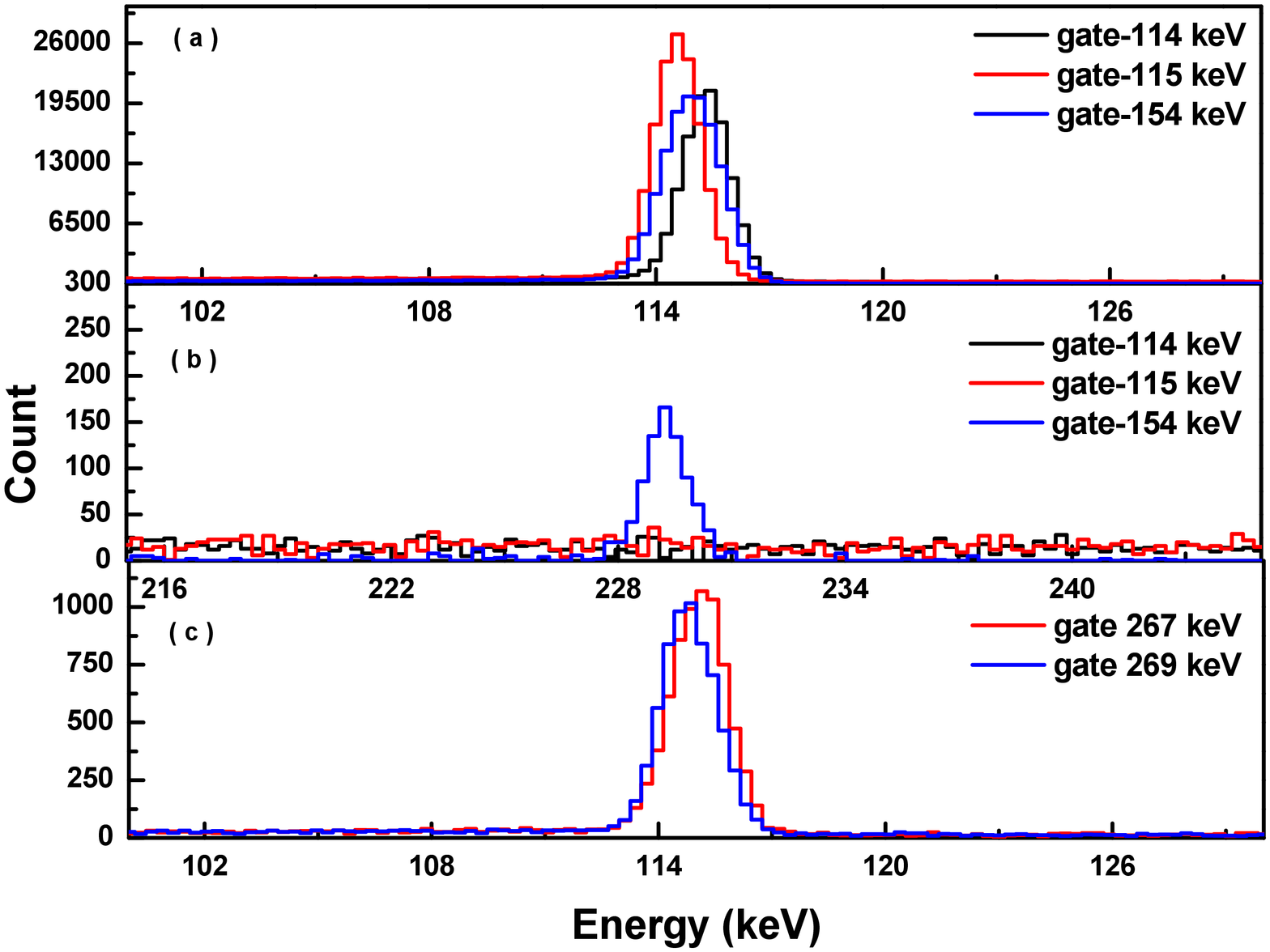}
  \vskip -2.0cm
  \caption{(colour online) The gated spectra obtained from the 10$\%$ HPGe detector by putting gates on 114.06 keV, 114.88 keV, 153.86 keV, 267.02 keV and 268.96 keV in LEPS detector are overlapped. Different energy regions of interest are shown in expanded scale.}
  \label{gate-overlap}
  \end{center}
\end{figure}

The singles $\gamma$ spectra acquired with the 50$\%$ HPGe detector, for a duration of 48h in each run and at definite time intervals, have been studied for the assignment of $\gamma$ rays in the level scheme of $^{146}$Eu. The counting was performed after allowing  a significant cooling time in order to exclude the contribution from the short-lived activities in the decay spectrum. All the $\gamma$ rays observed in the total spectrum are shown in FIG.~\ref{total-decay} and the $\gamma$ rays, relevant to the level scheme of $^{146}$Eu, were studied for the decay half lives. These data were also used for the measurement of energy and intensity of the transitions. The $\gamma$ rays observed in the total spectrum can be classified in four categories, viz., (i) the $\gamma$ rays from the decay of $^{146}$Gd (48d), (ii) $\gamma$ rays produced from the decay of $^{146}$Eu (5d), (iii) $\gamma$ rays from the decay of $^{147}$Eu(24d) and $^{149}$Gd (9.4d) produced in the reaction and (iv) the background and few unidentified $\gamma$ rays. The $\gamma$ rays of latter three categories were not considered in the present work. The decay plots for the relevant $\gamma$ rays are shown in FIG.~\ref{decay-all}.
The coincidence information has been derived from the data taken with the Ge-setup using $^{146}$Gd activity in acid solution as discussed in section~\ref{expt}. The gates were placed on 114.06, 114.88, 153.86, 267.02 and 268.96 keV transitions in the spectrum of LEPS detector (shown in FIG.~\ref{leps-seg1}) and the projected spectra obtained in the 10$\%$ HPGe detector are shown for different combinations in FIG.~\ref{gate-all} and ~\ref{gate-overlap}.
In the present work, only 114.06, 114.88, 153.86, 267.02 and 268.96 keV transitions were observed both in the coincidence as well as singles data. It is observed that 114.06 keV and 114.88 keV transitions are in coincidence with one another and also in coincidence with the 153.86 keV transition. The 229.4 keV transition is observed only in the gate of 153.86 keV but not in the gates of 114.06 and 114.88 keV transitions. The 268.96-keV transition has been observed only in the gate of 114.06 keV but not in 114.88 and 153.86 keV gates. The overlapped spectra of 267.06 and 268.96 keV transitions show that only 114.06 keV is present in the 268.96 keV gate whereas both the 114.06 and 114.88 keV transitions are present in the 267.06 keV gate. The aforesaid observation is also supported with the fact that the FWHM value for the peak obtained in the 267.06 keV gate is higher than that obtained in 268.96 keV gate. From these observations, the placements of the 114.06, 114.88, 153.86 and 267.06 keV gamma rays were confirmed.
The measured half lives of 229.4 and 268.86 keV transitions in the present work rule out the possibility of these two transitions to be originated from the summing effect. Hence, the 229.4 keV and 268.86 keV transitions were included parallel to the 114.06-114.88 keV and 114.88-153.86 keV cascades respectively, as evident from their coincidence data and decay half lives explained above.
The 420.88 and 575.24 keV $\gamma$ rays were observed in the singles data and their half lives confirm their placement in the level scheme. There was no indication of the 76 keV and 383 keV $\gamma$ rays, either in the coincidence data or in the singles data, and hence these two transitions were not placed in the present level scheme.
The 84.01, 276.57 and 627.35 keV gamma rays have been observed in the singles data of the present work and show the half lives of $\sim$48d. The first two transitions, viz. 84.01 and 276.57 keV, are reported to be produced from the decay of the 235 $\mu$s isomer as well as from the heavy ion fusion evaporation work~\cite{b10}. Again these two $\gamma$ rays are reported to be present in the 14-276-84-276 keV cascade according to the adopted level scheme of the nucleus. Between the two placements of the 276 keV transition, as decided by the above cascade, the one feeding the 84 keV $\gamma$ ray is reported to be decaying from the 647.5 keV level. However, the strongest transition decaying from this 647.5 keV level is 358 keV which was not found to exist in our data. Hence, the second 276 keV transition is not originated from the EC decay of $^{146}$Gd and is not considered in the present decay scheme.
Again, the calculated intensity of the 627.35 keV transition comes out to be equal to that of 276.57 keV transition. This observation suggests that the 627.35 keV transition is the one which is reported as 624.5 keV~\cite{nndc} decaying from the 914 keV level (modified as 918.5 keV level in the present work). Hence, it is concluded that the 918.5 keV level is produced from the EC decay of $^{146}$Gd and then feeds the 14.57-276.57-627.35 keV cascade.
A very high value of intensity ($\sim$ 12$\%$) for the 84.01 keV transition does not support it to be feeding the 276 keV $\gamma$ ray which subsequently feeds the 14 keV level. This observation, along with the similar intensity of 627.35 and 276.57 keV transitions as explained above, does not incorporate the 84.01 keV transition in the decay scheme. Based on all these information, the decay scheme of $^{146}$Gd has been modified considerably and shown in Fig. 6.
\begin{figure*}
  \begin{center}
  \vskip -1.0cm
  \hskip -6.0cm
  % Requires \usepackage{graphicx}
  \includegraphics[height=11.5cm, width=16cm,angle=-90]{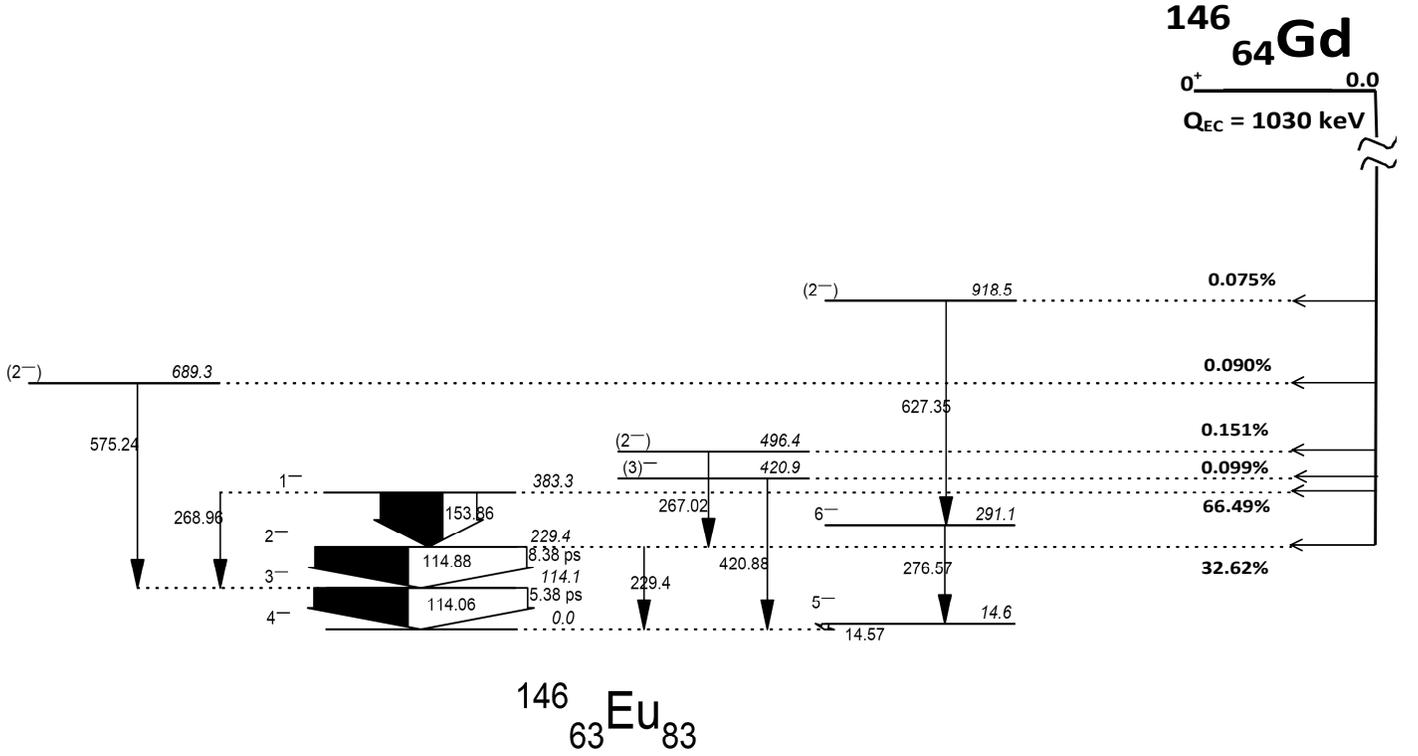}
  \vskip -4.5cm
  \caption{The level scheme of $^{146}$Eu from the decay of $^{146}$Gd as obtained from the present work. }
  \label{fig-levelscheme}
  \end{center}
\end{figure*}
 The energy and absolute intensity of the $\gamma$ rays belonging to $^{146}$Eu have been calculated from the singles measurement with 50$\%$ HPGe detector and indicated in TABLE~\ref{levelscheme}. While calculating the intensities, it has been considered that the intensities of the 14.57, 114.06, 229.4 and 420.88 keV transitions, feeding the ground state of $^{146}$Eu, add up to 100$\%$. Due to the fact that the 114.06 and 114.88 keV peaks could not be separated, the combined intensity of these two transitions was calculated from the total area obtained in the singles data. This total intensity has been distributed among these two energies according to the following equations by considering that no direct EC decay is feeding the 114.06 keV level.
 \begin{equation}\label{}
 \nonumber
   I_{114} + I_{115} = I_{total}
   \nonumber
 \end{equation}
 and
 \begin{equation}\label{}
 \nonumber
   I_{114} = I_{115} + I_{575} + I_{269}
 \end{equation}
The ratio of intensity for 114.06 keV to  that for 114.88 keV transition has also been verified from the total spectrum obtained in one segment of the LEPS detector as the two peaks could be separated with this detector.  The intensity of the 14.57 keV transition has been considered to be equal to that of 627.35 keV as discussed above. However, the intensity of this transition is much smaller than that of 114.06 keV and lies within the error limit. The normalised intensities have been used in order to develop the decay scheme of $^{146}$Gd ground state to different levels of $^{146}$Eu as shown in Fig.~\ref{fig-levelscheme}.
The logft values have been calculated from the LogFT calculator of NNDC~\cite{nndc-1} for different EC decays and are shown in TABLE~\ref{levelscheme}. These values are in accordance with the existing assignment of J$^{\pi}$ values for the states populated in $^{146}$Gd decay. The spin parity of all the states was thus kept unchanged with respect to the adopted decay scheme~\cite{nndc} except the 918.5 keV level which has been assigned to a J$^{\pi}$ value of 2$^-$.
The obtained half lives, relevant to identify the $\gamma$ rays involved in the decay of $^{146}$Gd, are also shown in the same TABLE~\ref{levelscheme}.
%\vskip -1.0cm
\begin{table*}
\caption{The populated levels of $^{146}$Eu from the EC decay of $^{146}$Gd.}
\begin{tabular}{|c|ccc|c|c|c|c|c|}
\hline\hline
     $E_i$   &$ J^{\pi}_i $&$ \rightarrow $&$ J^{\pi}_f $&$ E_{\gamma} $$^{\footnotemark[1]}$&$ Intensity ($\%$) $$^{\footnotemark[2]}$&half life(d)&Ec feeding ($\%$)&Logft$^{\footnotemark[3]}$\\
\hline

   14.6&$   5^{-} $&$ \rightarrow $&$   4^{-} $&  14.6& 0.075(11)& - &-&-\\
   114.1 &$   3^{-} $&$ \rightarrow $&$   4^{-} $&  114.06$\pm$0.05& 99.64 $\pm$3.21& 48.30$\pm$0.31$^{\footnotemark[4]}$&-&\\

   229.4 &$   2^{-} $&$ \rightarrow $&$   3^{-} $&  114.88$\pm$0.05& 98.74 $\pm$3.18&48.30$\pm$0.31&32.63&7.8\\

   383.3 &$   1^{-} $&$ \rightarrow $&$   2^{-} $&  153.86$\pm$0.01& 66.16 $\pm$1.02&48.94$\pm$0.64&66.52&7.3\\

   229.4 &$   2^{-} $&$ \rightarrow $&$   4^{-} $&  229.40$\pm$0.03& 0.188 $\pm$0.009&45.83$\pm$5.60&32.63&7.8\\

   496.4 &$   (2^{-}) $&$ \rightarrow $&$   2^{-} $&  267.02$\pm$0.08& 0.151 $\pm$0.029&45.11$\pm$9.16&0.151&9.7\\

   383.3 &$ 1^{-} $&$ \rightarrow $&$   3^{-} $&  268.96$\pm$0.06& 0.335 $\pm$0.027&53.47$\pm$5.15&66.52&7.3\\

   291.1 &$  6^{-}  $&$ \rightarrow $&$   5^{-} $&  276.57$\pm$0.08& 0.067 $\pm$0.007&42.46$\pm$9.99&-&-\\

   420.9 &$   (3)^{-} $&$ \rightarrow $&$   4^{-} $&  420.88$\pm$0.09& 0.099 $\pm$0.011&40.63$\pm$6.30&0.099&10\\

   689.3 &$  (2^-)  $&$ \rightarrow $&$   3^{-} $&  575.24$\pm$0.05& 0.090 $\pm$0.008&45.12$\pm$9.16&0.090&9.5\\

   918.5 &$ (2^-)  $&$ \rightarrow $&$   6^{-} $&  627.35$\pm$0.07& 0.075$\pm$0.011&64.16$\pm$2.99&0.075&8.4\\

   \hline\hline

\end{tabular}
\label{levelscheme}
 \footnotetext[1]{Only statistical errors in $\gamma$ energies are considered.}
\footnotetext[2]{The errors in the efficiencies and area under the photopeaks were considered for calculating the error in the intensities. For low energy transitions conversions coefficients were considered. The values of the coefficients were obtained for the Bricc calculator of NNDC.}
\footnotetext[3]{The Logft values were calculated from the LogFT calculator of NNDC.}
\footnotetext[4]{As the 114 and 115 keV peaks could not be separated in the spectrum obtained with the 50$\%$ HPGe detector, both the peaks were considered while calculating the half life.}
\end{table*}

\subsection{Lifetime Measurement}
\label{life}
The lifetime of the excited levels of $^{146}$Eu was measured following the mirror symmetric centroid difference (MSCD) method proposed by Regis et al~\cite{regis}. The conventional centroid shift method~\cite{mach1,mach2} is limited by the fact that the prompt reference curve depends on the gamma ray energy.  In case, the prompt references are not known or can not be extended for the gamma rays of interest with the standard sources, the Compton-Compton events are used for generating the prompt curves. Because of the inherent delay between a Compton-Compton and a photopeak-photopeak coincidence event, added up with other spurious events, such prompt curves are shifted significantly from the true prompt line and affect the measurements of lifetimes $\sim$ few ps~\cite{mach2}.  Whereas, in MSCD method, the lifetime can be derived as a function of the energy difference between the two gamma rays which are feeding to and decaying from the level of interest. Due to this fact, the MSCD technique allows the use of both the photopeak-photopeak and Compton-Compton events for the generation of prompt curves.
Following the conventional centroid shift method the lifetime $\tau$ of a specific branch is given by,
\begin{small}
\begin{eqnarray}
% \nonumber to remove numbering (before each equation)
%\nonumber
  \tau =C(D)_{stop} - C(P)_{stop}
  = - C(D)_{start} +  C(P)_{start}
\end{eqnarray}
\end{small}
 where C(D)$_{start}$ and C(D)$_{stop}$ are the TAC centroids of the delayed coincidence when references are taken from start detector and stop detector respectively. C(P)$_{start}$ and C(P)$_{stop}$ are similar TAC centroids for the prompt coincidences obtained from a prompt reference source.
 The C(P)$_{start}$ (or C(P)$_{stop}$), obtained by varying the $\gamma$ energy gates in the stop(or start)detector, generates the prompt reference curve when plotted as a function of those $\gamma$ energy values. The corresponding centroid of the level of interest ( C(D)$_{start}$ or C(D)$_{stop}$ ) is shifted from the prompt reference curve by the level lifetime ($\tau$). Whereas, the MSCD technique modifies this centroid shift method in the following way:
\begin{eqnarray}
\nonumber
\Delta C(D)& = & C(D)_{stop} - C(D)_{start}\\
\nonumber
& = & \tau+C(P)_{stop}+\tau-C(P)_{start}\\
\nonumber
& = & 2\tau + C(P)_{stop} - C(P)_{start}\\
& = & 2\tau + \Delta C(P)
\label{mscd}
\end{eqnarray}

 In this technique, the Prompt Reference Distribution (PRD) is generated from the plot of $\Delta$C(P) against the $\Delta E_{\gamma}$ (difference in the $\gamma$ energies between one feeding to and that decaying from a prompt level) using a prompt source, as explained in ref.~\cite{regis}. For a particular value of $\Delta E_{\gamma}$, the $\Delta$C(D), obtained from a delayed source, is shifted by twice the lifetime of the level of interest (2$\tau$) from the PRD curve as explained in the equation~\ref{mscd}.

\begin{figure}
  \begin{center}
  \vskip 0.3cm
  \hskip -2.5cm
  % Requires \usepackage{graphicx}
  \includegraphics[height=0.8\columnwidth, angle=-90]{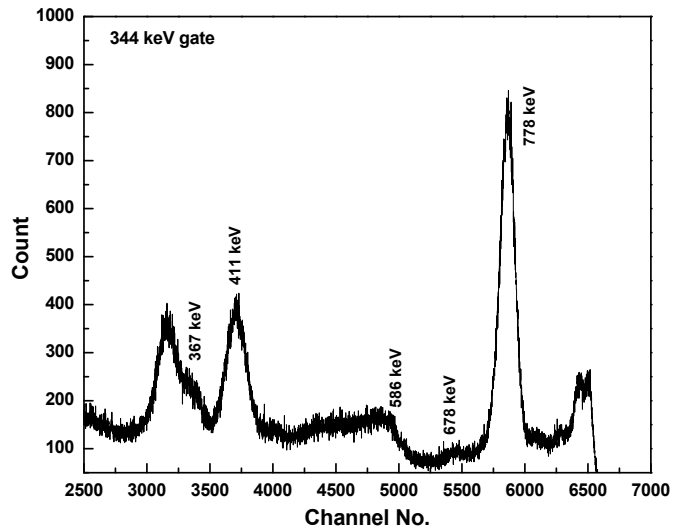}
  \vskip -2.0cm
  \caption{The $\gamma$ spectrum obtained with the `START' LaBr$_3$(Ce) detector when a 344 keV gate is placed in the 180$^o$ `STOP' detector.}
  \label{energy-eu}
  \end{center}
\end{figure}

\begin{figure}
  \begin{center}
  \vskip -0.25cm
  \hskip -1.3cm
  % Requires \usepackage{graphicx}
  \includegraphics[height=0.7\columnwidth, angle=-90]{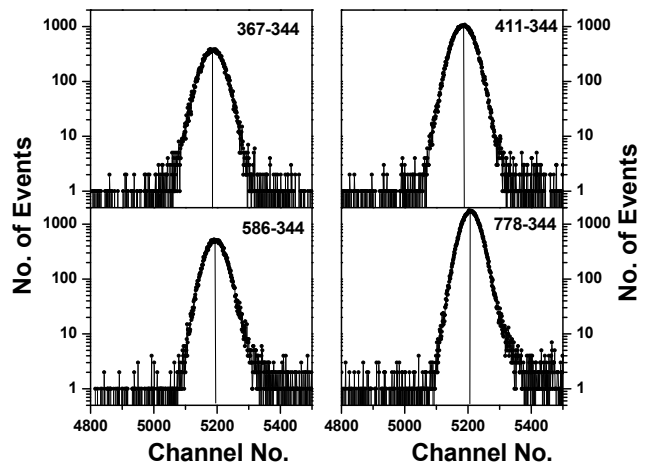}
  \vskip -1.8cm
  \caption{The TAC spectra obtained with 344 keV `STOP' gate in 180$^o$ detector and different `START' gates with $\gamma$ transitions feeding the 344 keV level of $^{152}$Gd. The increase in the TAC centroid are clearly visible.}
  \label{tac-eu1}
  \end{center}
\end{figure}

\begin{figure}
  \begin{center}
  \vskip -1.0cm
  \hskip -2.0cm
  % Requires \usepackage{graphicx}
  \includegraphics[height=0.5\columnwidth, width=9cm, angle=-90]{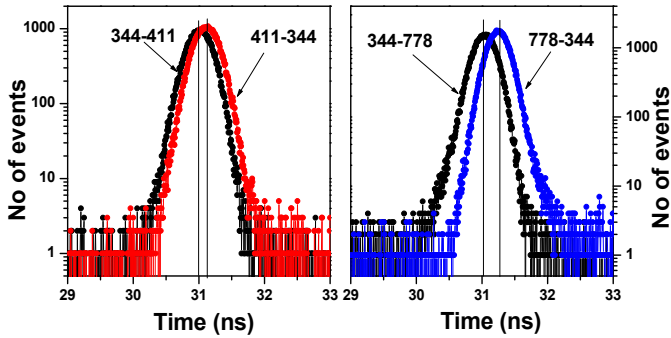}
  \vskip -3.1cm
  \caption{(colour online) TAC spectra obtained by interchanging `START' and `STOP' gates for 411-344 and 778-344 keV cascades of $^{152}$Gd are overlapped.}
  \label{tac-eu2}
  \end{center}
\end{figure}

\begin{figure}
  \begin{center}
  \vskip 0.25cm
  \hskip -2.5cm
  % Requires \usepackage{graphicx}
  \includegraphics[height=0.8\columnwidth, width=9cm, angle=-90]{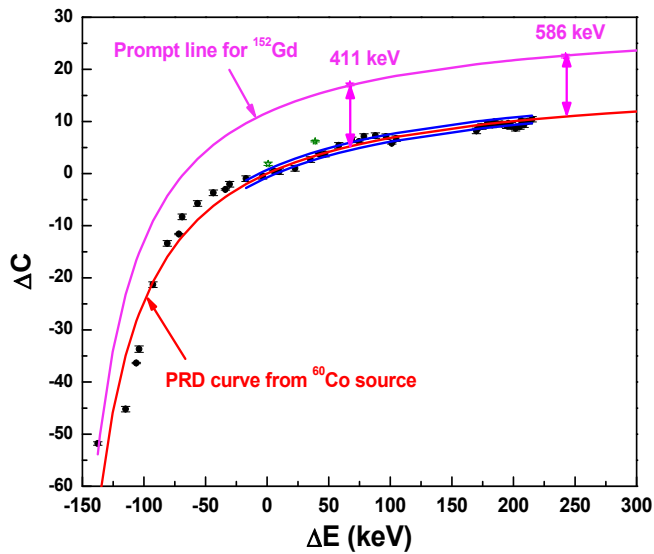}
  \vskip -1.8cm
  \caption{(colour online) PRD curve generated by fitting data points of $^{60}$Co (black circles) is shown with a solid red line. Two solid blue lines show upper and lower error limits of prompt curve. Data points for 344 keV level of $^{152}$Gd are shown by pink $\star$ s. Pink line approximately shows a shift of 35ps from PRD curve.}
  \label{prd}
  \end{center}
\end{figure}

\begin{figure}
  \begin{center}
  \vskip 0.3cm
  \hskip -2.8cm
  % Requires \usepackage{graphicx}
  \includegraphics[height=0.8\columnwidth, width=9cm, angle=-90]{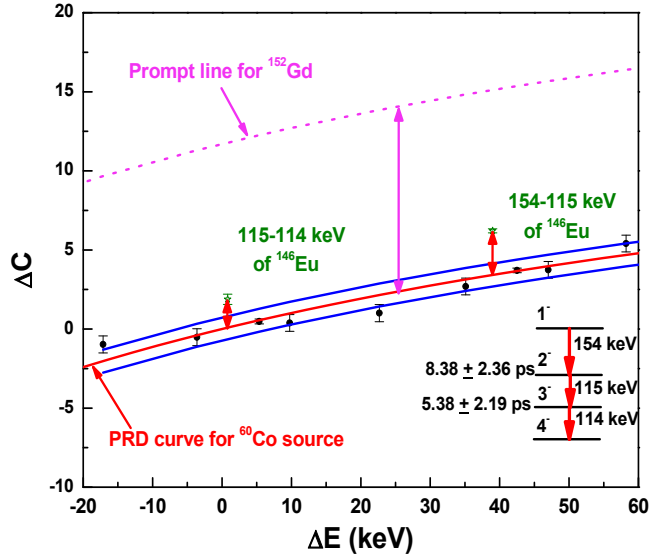}
  \vskip -1.6cm
  \caption{(colour online) A portion of FIG.~\ref{prd} is expanded for a closer view. The shift in $\Delta$C from red PRD line for the 115-114 and 154-115 keV cascades of $^{146}$Eu is clearly visible.}
  \label{prd-eu}
  \end{center}
\end{figure}

  In this work the LaBr$_3$-setup was used for the measurement of lifetimes of first 3$^-$ and 2$^-$ levels of $^{146}$Eu. Temperature stability was maintained in the experimental area to ensure a steady timing electronics. The $^{60}$Co prompt source was used to generate the PRD curve and the accuracy of our experimental setup was established with the measurement of 32.4 ps lifetime for the 344.2 keV level of $^{152}$Gd produced from the $\beta ^-$ decay of $^{152}$Eu source. The 344.2 keV level decays by a 344.2 keV $\gamma$ ray to the ground state of $^{152}$Gd and is fed by several $\gamma$ rays as known from its adopted level scheme~\cite{nndc-gd152}. The PRD curve was generated by using the Compton profile of the $^{60}$Co source with the reference energy gate at 344 keV in the `START' (`STOP') detector. In the `STOP' (`START') detector, a $\sim$10 keV wide energy gate was selected in the Compton profile of $^{60}$Co source at 10 keV intervals.
    The different coincident  photopeaks of $^{152}$Gd have been projected (shown in FIG.~\ref{energy-eu}) in the spectrum of `START' detector by putting gate at 344 keV in the 180$^{\circ}$ `STOP' detector. Due to the desired energy resolution of the LaBr$_3$(Ce) detector, all the peaks could be clearly identified. The TAC spectra were obtained with the gate at 344 keV photopeak in the 180$^{\circ}$ `STOP' detector and the gates at the other photopeaks, in coincidence with 344 keV, given in the `START' detector. Some of these TAC spectra are shown in FIG.~\ref{tac-eu1} in order to envisage the dependence of the TAC centroid on the $\gamma$ energy. Similar TAC spectra were obtained with the 344 keV gate at the `START' detector and other photopeaks at the `STOP' detector. The difference in the TAC centroids obtained by interchanging the `START' and `STOP' energy gates is shown in FIG.~\ref{tac-eu2} with a special reference to the 411-344 and 778-344 keV cascades. The similar differences in the TAC centroid were also obtained for the other cascades of $^{152}$Gd and have been utilized for the measurements of the lifetime of 344.2 keV level.
The PRD curve, generated with the $^{60}$Co source, is shown in FIG.~\ref{prd} and fitted with a function $f(\Delta E_{\gamma}) = \frac{a\Delta E_\gamma}{b+\Delta E_{\gamma}}$, where a and b comes out to be 18.96$\pm$0.145 and 177.3 $\pm$ 2.39 respectively.
The fitted curve is shown with red solid line. The two blue lines above and below the red PRD line represent the positive and negative error limits of the PRD curve respectively. The limits have been considered to be the standard deviation of the experimental data points from the fitted PRD line. It is observed that the prompt curve can be determined within a limit of 4.35 ps on an average for the positive $\Delta$E values.
The $\Delta$C values obtained for 411-344 and 586-344 cascades of $^{152}$Gd are also shown in the same figure. The shift of these data points from the PRD curve gives $\tau$ = 35.26 $\pm$ 2.75 ps, which is in well agreement with the literature value ( 32.4 $\pm$ 1.7 ps )~\cite{nndc-gd152} for the lifetime of 344 keV level of $^{152}$Gd.
%However, the similar shifts for the 658-344 and 778-344 cascades show a higher value ( $\sim$ 60 ps ) for the lifetime of this 344 keV level. This is probably because of the non linearity of the LaBr$_3$ (Ce)detector affecting the calibration at higher $\Delta$E values under the present working voltages of the photomultiplier tube.
After the lifetime for the 344 keV level of $^{152}$Gd was reproduced satisfactorily, the similar centroid differences were also obtained for the level of $^{146}$Eu and shown in Fig.~\ref{prd}. The 114.06-114.88 and 114.88-153.86 keV cascades were used for deducing the lifetimes of first 3$^-$ and 2$^-$ states of $^{146}$Eu respectively.
  In FIG.~\ref{prd-eu}, the $\Delta C$ values obtained for the $^{146}$Eu are shown with respect to the PRD curve of FIG.~\ref{prd}, in an expanded scale. The centroid differences and the corresponding lifetimes for the relevant cascades of $^{152}$Gd and $^{146}$Eu are furnished in TABLE~\ref{timing}. The lifetime for 344 keV level of $^{152}$Gd measured in the present work matches satisfactorily with the value obtained from the latest compilation in Nuclear Data Sheets~\cite{nndc-gd152}. In case of first 3$^-$ and 2$^-$ states of $^{146}$Eu, the present study removes the limits~\cite{nndc} by assigning lifetime values within an acceptable error.

%\vskip -1.3cm
\begin{table}
\caption{The lifetime for first two excited levels of $^{146}$Eu and 344 keV level of $^{152}$Gd, obtained from the present work.}
\begin{tabular}{|c|c|c|c|c|c|}
\hline\hline
Nucleus&Cascade&$\Delta$C&PRD&$\tau$ (ps)&Lit. value\\
&&&&(present work)&\\
\hline
&&&&&\\
$^{152}$Gd&(411-344)&17.11&5.21&35.71$\pm$2.29&32.4$\pm$1.7~\cite{nndc-gd152}\\
&&&&&\\
&(586-344)&22.54&10.94&34.78$\pm$2.40&\\
&&&&&\\
$^{146}$Eu&(115-114)&1.88&0.087&5.38$\pm$2.36&$\le$0.16ns~\cite{nndc}\\
&&&&&\\
&(154-115)&6.21&3.42&8.38$\pm$2.19&$\le$0.2ns~\cite{nndc}\\
\hline\hline
\end{tabular}
\label{timing}
\end{table}

\section{Shell Model Calculation}
%\subsection{Shell model calculation}
\label{sm}
A large basis shell model calculation was performed using the code OXBASH \cite{BAB94}
in order to characterize the low-lying states of $^{146}$Eu. The calculation considered
$^{132}$Sn as core and thirteen protons distributed over the model space
comprising of $\pi$(1$g_{\frac{7}{2}}$, 2$d_{\frac{5}{2}}$, 2$d_{\frac{3}{2}}$, 3$s_{\frac{1}{2}}$)
single particle orbitals along with one neutron over the $\nu$1$f_{\frac{7}{2}}$
single particle orbital. The calculations were carried out using
proton-neutron formalism in full particle space. The two-body matrix
elements were obtained from the well-known $\it{n82pota}$ interaction
supplied with the source code of OXBASH. This interaction
is basically a combination of $\it{rotann}$, $\it{rotapp}$ and $\it{rotapn}$ interaction
files which are also supplied with the code.
The $\it{rotapp}$ TBMEs were obtained by Brown et al.~\cite{brown88} by a 5-parameter
fit to $H7B$ interaction~\cite{hosaka}  potential for
N=82 core and the corresponding NN TBMEs were obtained in the same way but
by excluding the Coulomb interaction. The bare $G$ matrix from the $\it{H7B}$ was used for
the PN TBMEs. In our calculations, we have truncated the neutron space
to contain only $1f_{7/2}$ orbital as well as the proton space by excluding
the $1h_{11/2}$.
In the $\it{n82pota}$ interaction, the neutron single particle energies (SPEs)
were chosen to give the -7.48 MeV energy difference betwen $^{147}$Gd and
$^{146}$Gd as well as to best reproduce the excitation energies of $\frac{7}{2}$, $\frac{9}{2}$
and $\frac{13}{2}$ states in $^{149}$Dy spectrum. The proton SPEs were obtained from
the potential fit as mentioned above.

TABLE~\ref{shmod} shows calculated excitation energies of the negative parity states
up to the $1^-$ level at 383.3 keV along with the theoretical
and experimental lifetimes and the major configuration of the states.
 The lifetime values have been determined from the reduced transition probabilities obtained from the calculation for different excited states of $^{146}$Eu. A remarkable agreement for the energies and lifetimes is noticeable.
It is also noticeable that the major contribution to the structure of the
states comes from the $\pi g_{\frac{7}{2}}$ and $\pi d_{\frac{5}{2}}$ orbitals which constitute
the $^{146}$Gd core ground state.\\
\begin{table}
\caption{The level energies and lifetimes are compared with shell model calculation. The major configurations responsible for the low lying levels of $^{146}$Eu are shown. }
\begin{tabular}{|c|cc|cc|c|}
\hline\hline
J$^{\pi}$&\multicolumn{2}{c|}{E$_{level}$}&\multicolumn{2}{c|}{Lifetime}&Configuration\\
%\cline{7-13}
&\multicolumn{2}{c|}{(keV)}&\multicolumn{2}{c|}{(ps)}&\\

&(expt)&(th.)&(expt)&(th.)&\\

\hline
    &&&&&\\

    $   4^{-} $&0.0&0.0&-&-&[$\pi(g_{\frac{7}{2}}^8, d_{\frac{5}{2}}^5) \nu(f_{\frac{7}{2}} )^1$]\\
&&&&&(82$\%$)\\
   &&&&&+ [$\pi(g_{\frac{7}{2}}^ 8, d_{\frac{5}{2}}^4, d_{\frac{3}{2}}^1) \nu(f_{\frac{7}{2}})^1$]\\
&&&&&(12.3$\%$)\\
    &&&&&+ [$\pi(g_{\frac{7}{2}}^8, d_{\frac{5}{2}}^4, s_{\frac{1}{2}}^1) \nu(f_{\frac{7}{2}})^1$]\\
&&&&&(4.1$\%$)\\

    &&&&&\\

    $   5^{-} $&14.57&16.0&-&-&[$\pi(g_{\frac{7}{2}}^7, d_{\frac{5}{2}}^6) \nu(f_{\frac{7}{2}} )^1$]\\
   &&&&&(77.3$\%$)\\
&&&&&+ [$\pi(g_{\frac{7}{2}}^ 7, d_{\frac{5}{2}}^4, d_{\frac{3}{2}}^2) \nu(f_{\frac{7}{2}})^1$]\\
&&&&&(15.2$\%$)\\

    &&&&&\\

   $   3^{-} $&114.06&115&5.38$\pm$2.36&5.4&[$\pi(g_{\frac{7}{2}}^7, d_{\frac{5}{2}}^6) \nu(f_{\frac{7}{2}} )^1$]\\
   &&&&&(62.7$\%$)\\
&&&&&+ [$\pi(g_{\frac{7}{2}}^ 8, d_{\frac{5}{2}}^5) \nu(f_{\frac{7}{2}})^1$]\\
&&&&&(20$\%$)\\
   &&&&&+ [$\pi(g_{\frac{7}{2}}^7, d_{\frac{5}{2}}^4, d_{\frac{3}{2}}^2) \nu(f_{\frac{7}{2}})^1$]\\
&&&&&(11.5$\%$)\\

 &&&&&\\

   $   2^{-} $&229.4&232&8.38$\pm$2.19&3.04&[$\pi(g_{\frac{7}{2}}^8, d_{\frac{5}{2}}^5) \nu(f_{\frac{7}{2}} )^1$]\\
   &&&&&(63.8$\%$)\\
&&&&&+ [$\pi(g_{\frac{7}{2}}^8, d_{\frac{5}{2}}^4, d_{\frac{3}{2}}^1) \nu(f_{\frac{7}{2}})^1$]\\
&&&&&(34.3$\%$)\\

 &&&&&\\

   $  6^{-}  $&290.6&276&-&-&[$\pi(g_{\frac{7}{2}}^8, d_{\frac{5}{2}}^5) \nu(f_{\frac{7}{2}} )^1$]\\
&&&&&(98.4$\%$)\\
 &&&&&\\

   $   1^{-} $&383.3&385&-&-&[$\pi(g_{\frac{7}{2}}^7, d_{\frac{5}{2}}^6) \nu(f_{\frac{7}{2}} )^1$]\\
    &&&&&(82$\%$)\\
&&&&&+ [$\pi(g_{\frac{7}{2}}^ 7, d_{\frac{5}{2}}^4, d_{\frac{3}{2}}^2) \nu(f_{\frac{7}{2}})^1$]\\
&&&&&(13.3$\%$)\\
    &&&&&+ [$\pi(g_{\frac{7}{2}}^7, d_{\frac{5}{2}}^5, d_{\frac{3}{2}}^1) \nu(f_{\frac{7}{2}})^1$]\\
&&&&&(10.7$\%$)\\

 &&&&&\\
\hline\hline
\end{tabular}

\label{shmod}
%\footnotetext[1]{The values have been calculated from the measured lifetimes.}
%\footnotetext[2]{The values have been calculated from the measured intensities.}
\end{table}

\section{Discussion}
\label{dis}

The decay spectroscopy of $^{146}$Gd nucleus has been performed using the light ion beams from Variable Energy Cyclotron Centre to study the low lying structure of the odd odd $^{146}$Eu nucleus. The low lying excited levels of this nucleus have been considerably modified from the singles decay and $\gamma$$\gamma$ coincidence measurements. The present study assigns new levels in the decay of $^{146}$Gd and confirms two crossover transitions of 229.4 and 268.96 keV in the level scheme of $^{146}$Eu.
The lifetimes of the first 3$^-$ and 2$^-$ levels of $^{146}$Eu have been measured to be 5.38 $\pm$ 2.36 and 8.38 $\pm$ 2.19 ps respectively following the Mirror Symmetric Centroid Difference technique using LaBr$_3$(Ce) detectors. Till date the literature values for the lifetimes of these two states provide only the limits while the present study assigns definite lifetime values with an acceptable error. A Shell model calculation has been performed using OXBASH code considering $^{132}$Sn as a core nucleus. The theoretical calculation agrees remarkably well with the experimental data. The level spectra and the lifetime values of the first two excited states along with the shell mode calculation establish the near spherical structure of $^{146}$Eu vis a vis the validity of Z = 64 subshell closure for N = 82 closed shell nuclei.\\

\section{Acknowledgement}
\label{ack}
The authors acknowledge the efforts of the operators of K=130 Cyclotron of Variable Energy Cyclotron Centre for providing a good quality beam. The valuable suggestions from Dr. S. K. Basu is gratefully acknowledged. One of the authors (Mr. T. Malik) is thankful to Head, Physics Group, VECC and Head of the Department, Department of Applied Physics, Indian School of Mines, for allowing him to carry his winter project in VECC during the period Decemeber 2012 to January 2013.  The authors acknowledge the sincere efforts of Mr. R. K. Chatterjee  who have assisted meaningfully in target preparation. The authors have been highly benefitted by the presence of Dr. H. Pai, Mr. P. Mukhopadhyay and Mr. A. Ganguly in maintaining Ge detectors during the experiment.

%\vskip 10.0cm

\end{document}